%% file: Xi0_with_BNS.tex.tex
\documentclass[11pt]{article}
\pdfoutput=1

\usepackage{jcappub}
\usepackage{mathrsfs,amssymb,amsmath,verbatim,mathtools,graphicx,bm,booktabs,tabularx,multirow,relsize,framed,orcidlink}
\usepackage{xcolor}
\usepackage[T1]{fontenc}
\usepackage[utf8]{inputenc}
\usepackage{tikz}

\usepackage[noabbrev]{cleveref} %for JCAP, use noabbrev, figure instead of Fig. etc
\crefname{section}{section}{sections}
\Crefname{section}{Section}{Sections}
\crefname{appendix}{appendix}{appendices}
\Crefname{appendix}{Appendix}{Appendices}
\crefname{footnote}{footnote}{footnotes}
\Crefname{footnote}{Footnote}{Footnotes}

\numberwithin{equation}{section}

\input{mydefs}

\newcommand{\unige}{D\'epartement de Physique Th\'eorique, Universit\'e de Gen\`eve, 24 quai Ernest Ansermet, 1211 Gen\`eve 4, Switzerland}
\newcommand{\gwsc}{Gravitational Wave Science Center (GWSC), Universit\'e de Gen\`eve, CH-1211 Geneva, Switzerland}
\newcommand{\mrs}{Aix-Marseille Universit\'e, Universit\'e de Toulon, CNRS, CPT, Marseille, France}
\newcommand{\ifae}{Institut de Física d’Altes Energies (IFAE), Barcelona Institute of Science and Technology, E-08193 Barcelona, Spain}

\graphicspath{figures/} 

\begin{document}

\title{Cosmology and modified GW propagation from the BNS mass function  at third-generation detector networks}

\author[a,1,2]{Dounia Nanadoumgar-Lacroze \orcidlink{0009-0009-7255-8111} \note{Authors to whom correspondence should be addressed.} \note{This work is part of the doctoral thesis of D.N.L. within the framework of the Doctoral Program in Physics at the Autonomous University of Barcelona.}}
\author[b,c,1]{Niccol\`o Muttoni \orcidlink{0000-0002-4214-2344}}
\author[b,c]{Michele Maggiore \orcidlink{0000-0001-7348-047X},}
\author[d]{Michele Mancarella \orcidlink{0000-0002-0675-508X}}
%\footnote{$\dagger$ }

\affiliation[a]{\ifae}
\affiliation[b]{\unige}
\affiliation[c]{\gwsc}
\affiliation[d]{\mrs}

\emailAdd{dnala@ifae.es}
\emailAdd{niccolo.muttoni@unige.ch}
\emailAdd{michele.maggiore@unige.ch}
\emailAdd{mancarella@cpt.univ-mrs.fr}

\abstract{We perform forecasts for the Hubble parameter $H_0$ and for the parameter $\Xi_0$ that describes modified gravitational-wave propagation,  using information from the binary neutron star (BNS) mass function, for  Einstein Telescope (ET), taken either in the triangle or in the ``2L'' configuration, as well as for detector network made by ET together with a 40-km Cosmic Explorer (CE). We  restrict  ourselves to BNSs  with a large signal-to-noise ratio, ${\rm SNR} >50$, which still give $\mathcal{O}(10^3)$ events yr$^{-1}$, and we perform a full joint cosmology-population Bayesian inference. 
We find that, for ET in isolation, the two ET configurations perform comparably, yielding uncertainties of 12\% and 11\% on $H_0$ for the triangular and 2L designs, respectively, and 18\% uncertainty on $\Xi_0$ in both cases. For networks including ET and CE, we can constrain $H_0$ and $\Xi_0$ to precisions of 9\% and 6\%, respectively. These results should be taken as a very conservative estimate of third-generation detectors' capabilities as a consequence of the high SNR cut. We project the constraints on the $\Lambda \rm CDM$ expansion history and find that ET  alone (triangular and 2L configurations) achieves its best precision on $H(z)$ at $z=0.23$ and $z=0.28$, yielding a 10\% and 6\% precision, respectively. When CE is added to the network, the precision improves to 4\% and 3\% at $z=0.37$ and $z=0.38$, respectively.}

\maketitle
\flushbottom

\section{Introduction}\label{intro}

The possibility of extracting cosmological information from  gravitational-wave (GW) observations of compact binary coalescences (CBCs) was recognized long ago ~\cite{Schutz:1986gp}, and has been the subject of many subsequent  studies, e.g.~\cite{Chernoff:1993th,Holz:2005df,Dalal:2006qt,MacLeod:2007jd,Nissanke:2009kt,Cutler:2009qv,DelPozzo:2011vcw,Taylor:2011fs,Taylor:2012db,Chen:2017rfc,Feeney:2018mkj,Gray:2019ksv,Mukherjee:2020hyn,Mastrogiovanni:2020mvm,Mukherjee:2020mha,Finke:2021aom,Finke:2021znb,Finke:2021eio,Palmese:2021mjm,Mancarella:2021ecn,Ezquiaga:2022zkx,Leyde:2022orh,Mukherjee:2022afz,Borghi:2023opd,Ferri:2024amc}. The cosmological usefulness of CBCs  is due to the fact that their GW signal  allows us to infer the luminosity distance to the source; for this reason these sources are called ``standard sirens'', by analogy with the ``standard candles'' that have a standard (or, rather, standardizable) electromagnetic luminosity. Since the information on cosmology is contained in the relation between luminosity distance and redshift,  an independent determination  of the redshift to the CBC is also necessary. The GW signal does not carry immediate information on the redshift because of the mass-redshift degeneracy, i.e. the masses $m_i$ of the two initial bodies and the redshift $z$ enter the waveform in the combination $m_i(1+z)$.  
The simplest scenario that allows us to obtain the source redshift takes place when the GW signal from a CBC has an electromagnetic counterpart (``bright sirens''), so we can identify the host galaxy from the electromagnetic signal. This, however, is now expected to be a rare occurrence, at least at the  current second-generation GW detectors LIGO, Virgo and KAGRA. Currently,  the only observation of this type is the celebrated binary neutron star (BNS) coalescence GW170817~\cite{TheLIGOScientific:2017qsa,LIGOScientific:2017zic}. 

In the absence of an electromagnetic counterpart (``dark sirens''), various statistical methods have been developed, such as the  combination of a large set of GW events with galaxy catalogs \cite{Schutz:1986gp,DelPozzo:2011vcw,Chen:2017rfc,Finke:2021aom,Gray:2021sew,Gair:2022zsa,Mastrogiovanni:2023zbw,Borghi:2023opd,Ferri:2024amc} or the correlation with features in the mass distribution of black holes and neutron stars (NSs)~\cite{Taylor:2011fs,Taylor:2012db,Farr:2019twy,Mastrogiovanni:2021wsd,Mancarella:2021ecn,Finke:2021eio,Ezquiaga:2022zkx,Chen:2024gdn,Pierra:2026ffj,Tagliazucchi:2026gxn} (``spectral sirens''); see \cite{Palmese:2025zku,Pierra:2025fgr} for recent reviews. Other possibilities to break the mass-redshift degeneracy include the observation of tidal effects and extraction of the Love number in BNS coalescences  (``Love sirens'', \cite{Messenger:2011gi}).

The perspectives for observations of both bright and dark sirens will improve dramatically at third-generation (3G) detectors, such as the Einstein Telescope (ET) \cite{Punturo:2010zz,Hild:2010id,Maggiore:2019uih} and Cosmic Explorer (CE) \cite{Reitze:2019iox,Evans:2021gyd,Evans:2023euw}. In particular, 3G detectors are expected to observe $\calO(10^5)$ binary black hole (BBH) and BNS events per year \cite{Branchesi:2023mws,ET:2025xjr} and, with such a large  statistical sample, dark sirens techniques can become very effective. Another scenario that opens up for 3G detectors is the possibility of localizing a GW event sufficiently well that in the localization volume there is just one galaxy \cite{Borhanian:2020vyr,Muttoni:2023prw}; for 3G detectors there can be a handful of such events per year, see figure~2.28 of  \cite{ET:2025xjr}.
We refer the reader to Section~2.3.1 of \cite{ET:2025xjr} for a  review of the various dark sirens methods, and their applicability to 3G detectors.
It is important to stress that the inference of the cosmological parameters with these statistical techniques is intertwined with the inference of the astrophysical population parameters, such as the redshift and mass distribution of the sources \cite{Mastrogiovanni:2021wsd, Gair:2022zsa, Moresco:2022phi, Mastrogiovanni:2023emh,Gray:2023wgj, Borghi:2023opd} (or, for the Love sirens method, the parameters describing the NS equation of state).

In this paper we discuss how to extract cosmological parameters, and especially those that describe modified GW propagation (due to modifications of gravity at cosmological scales), using the BNS mass function. The idea of using the BNS mass function for extracting cosmological information was proposed in \cite{Chernoff:1993th,Taylor:2011fs,Taylor:2012db} in the context of the determination of $H_0$ within $\Lambda$CDM. In \cite{Finke:2021eio} it was discussed how to apply this idea to modified GW propagation, showing that the method can be quite promising for testing modifications of gravity. Here we will present a quantitative discussion of the perspectives at 3G detector networks, performing a joint cosmology and population Bayesian  analysis.
The paper is organized as follows. In section~\ref{sect:modGWprop}, we briefly review modified GW propagation and the current limits on the main parameter, $\Xi_0$, that describes it phenomenologically. In section~\ref{sect:setup}, we present the basic set-up of our analysis, i.e. the 3G detector networks used, the BNS population model, and the technical tools used in this study. In section~\ref{sect:results}, we present our results before concluding in section \ref{sect:conclusions}.

\section{Modified GW propagation}\label{sect:modGWprop}

The phenomenon of modified GW propagation refers to the fact that, in theories where General Relativity (GR) is modified at cosmological scales, the propagation of GWs across cosmological distances is, in general, also modified.  In particular, in  modified  gravity theories that do not affect the speed of GWs (which, after the observation of GW170817, is  now excluded at a level of about a part in $10^{15}$  \cite{LIGOScientific:2017zic}), the evolution of   tensor perturbations over a Friedmann-Lemaître-Robertson-Walker (FLRW) background is governed 
by the equation\footnote{We use standard notation: $\tilde{h}_A(\eta, {\bf k})$ is the Fourier-transformed GW amplitude,  $A=+,\times$ labels the two GW polarizations, the prime denotes the derivative with respect to cosmic time $\eta$, defined by $d\eta=dt/a(t)$, $a(\eta)$ is the scale factor, and 
${\cal H}=a'/a$.}
\begin{equation}\label{prophmodgrav}
\tilde{h}''_A  +2 {\cal H}[1-\delta(\eta)] \tilde{h}'_A+c^2k^2\tilde{h}_A=0\, ,
\end{equation}
where  $\delta(\eta)$ is a function that encodes the modifications from GR, so that GR is recovered for $\delta(\eta)=0$. This behavior has been  found in many explicit modified gravity models~\cite{Saltas:2014dha,Lombriser:2015sxa,Nishizawa:2017nef,Arai:2017hxj,Belgacem:2017ihm,Amendola:2017ovw,Belgacem:2018lbp,Nishizawa:2019rra,Belgacem:2019lwx,Belgacem:2020pdz,LISACosmologyWorkingGroup:2019mwx}.
A consequence of this modification is  that the amplitude of the GW generated by a CBC is no longer proportional to the inverse of the  luminosity distance $d_L(z)$ of the source (which, in this context, we denote by $\dem(z)$, since this is the quantity that would be measured from an electromagnetic counterpart). Rather, it is proportional to the inverse of a ``GW luminosity distance'' $\dgw(z)$ \cite{Belgacem:2017ihm}, related to $\dem(z)$ by~\cite{Belgacem:2017ihm,Belgacem:2018lbp}
\begin{equation}\label{dLgwdLem}
\frac{\dgw(z)}{\dem(z)}=\Xi(z)\, ,
\end{equation}
where 
\begin{equation}\label{Xiz}
\Xi(z)=\exp\left\{-\int_0^z \,\frac{dz'}{1+z'}\,\delta(z')\right\}\, ,
\end{equation}
and 
the function $\delta$ that appears in \eq{prophmodgrav} has now been written as a function of redshift. 
A full reconstruction of the function $\Xi(z)$ from future GW observations is a challenging task. Although partial reconstructions are possible, for instance using Gaussian process techniques~\cite{Belgacem:2019zzu}, it is often more practical, as in the case of the dark energy (DE) equation of state,  to work within a parametrized framework.  
A  convenient parametrization is given by~\cite{Belgacem:2018lbp}
\be\label{eq:fit}
\Xi(z)=\Xi_0 +\frac{1-\Xi_0}{(1+z)^n}\, ,
\ee
and is characterized by two parameters $(\Xi_0,n)$.
This form yields a smooth power-law interpolation in the scale factor $a=1/(1+z)$, connecting the present-day limit $d_L^{\,\rm gw}/d_L^{\,\rm em}=1$ at $z=0$ (corresponding to the absence of propagation effects) to a constant value at high redshift. The latter behavior is well motivated, since in typical modified gravity scenarios departures from GR occur only in the recent cosmological past. Consequently, $\delta(z)$ vanishes at large redshift and the integral in \eq{Xiz} approaches a constant. The GR limit is recovered for $\Xi_0=1$, independently of the value of $n$.
This simple parametrization has been shown to provide an excellent description covering a broad range  of well-studied modified gravity models such as $f(R)$ gravity,
\cite{Starobinsky:2007hu,Hu:2007nk}, Jordan-Brans-Dicke theory \cite{Brans:1961sx}, Galileon theories~\cite{Nicolis:2008in,Chow:2009fm}, and non-local gravity~\cite{Maggiore:2013mea,Maggiore:2014sia,Belgacem:2020pdz};
see  \cite{LISACosmologyWorkingGroup:2019mwx} for detailed discussion.
It should be observed that, in the parametrization (\ref{eq:fit}), GR is recovered not only for $\Xi_0=1$ (and arbitrary $n$) but also for $n=0$ (and arbitrary $\Xi_0$). To get rid of this potential ambiguity, when performing a Bayesian analysis we will set a prior $n\geq n_{\rm min}$, with $n_{\rm min}$ strictly positive, see section~\ref{sect:setup}.

Several limits on modified GW propagation have already been obtained from  current  GW observations (see section~2.3.2.2 of \cite{ET:2025xjr} for a recent overview). In general, significant limits on $n$ in \eq{eq:fit} cannot yet be obtained  from current data; however, interesting limits on deviations from GR, as quantified by the deviation of $\Xi_0$ from the GR value $\Xi_0=1$,
have already been found.
The most recent and stringent result, $\Xi_0 = 1.0^{+0.4}_{-0.2}$  (with a narrow prior on $H_0$, and $68\%$ c.l.), was obtained by the LVK Collaboration \cite{LIGOScientific:2025jau} using the GWTC-4.0 catalog of GW events~\citep{LIGOScientific:2025hdt,LIGOScientific:2025slb} and combining them with both galaxy catalogs and the mass distribution of CBCs. Related analyses based on earlier GW catalogs and different modeling assumptions were presented in \cite{Finke:2021aom, LIGOScientific:2020ibl, LIGOScientific:2021usb, KAGRA:2021vkt, Mancarella:2021ecn, Mastrogiovanni:2023emh, Leyde:2022orh, Chen:2023wpj}.

\section{Methodology}\label{sect:setup}

\subsection{Population and cosmology hyper-parameters}

To perform a joint inference on the population and cosmological parameters we use a hierarchical Bayesian approach. This technique is by now standard in this context, and we refer the reader to the original papers and reviews \cite{Loredo:2004nn, Adams:2012qw, Mandel:2018mve, Thrane:2018qnx, Vitale:2020aaz} for general discussion of the method.
We adopt the Bayesian framework developed in \textsc{Icarogw}, which models GW detections as a noisy, inhomogeneous and incomplete Poisson process \cite{Mastrogiovanni:2023zbw}. The hierarchical likelihood used to infer hyper-parameters $\Lambda$ is defined as 
\begin{equation}
    \mathcal{L}(\{x\}|\Lambda) \propto \prod_{i=1}^{N_{\rm obs}} \frac{\displaystyle \int\mathcal{L}(x_i|\theta,\Lambda) \frac{dN}{dtd\theta}(\Lambda) dt d\theta}{\displaystyle \int p_{\rm det}(\theta,\Lambda) \frac{dN}{dtd\theta}(\Lambda) dt d\theta},
\end{equation}
where $N_{\rm obs}$ is the number of detections, each described by a set of parameters $\theta$, in a dataset $\{x\}$; $\mathcal{L}(x_i|\theta,\Lambda)$ are the individual likelihood terms, $\frac{dN}{dtd\theta}(\Lambda)$ is the event production rate, and $p_{\rm det}(\theta,\Lambda)$ accounts for the selection effects. Integrals are estimated relying on Monte Carlo integration. More details can be found in \cite{Mastrogiovanni:2023zbw}.

We apply this method to a set of BNS observations, using as prior information the mass and redshift distributions. We do not combine it  with information from the galaxy catalog method (as in \cite{LIGOScientific:2025jau}), since our aim here is to assess the constraining power of the method based on the BNS mass function prior. The two crucial ingredients of the method  based on the mass and redshift information are the redshift prior and the mass prior. For the former, we take a Madau-Dickinson parametrization ~\cite{Madau:2014bja,Madau:2016jbv}, 
\be
\psi(z) \propto \dfrac{(1+z)^\gamma}{1 + \left(\dfrac{1+z}{1+z_p}\right)^{\gamma + \kappa}} \, ,
\ee
convoluted by a log-flat time-delay distribution to keep into account binary evolution (see analogous methodologies in refs.~\cite{Iacovelli:2022bbs,Mancarella:2024qle,Muttoni:2023prw}). The detector-frame BNS redshift distribution then reads
\be\label{Madau}
R(z \mid \Lambda_{z}) \propto \psi(z) \dfrac{1}{(1+z)} \dfrac{dV}{dz} \, ,
%R(z|\Lambda_z) = R_0 \,  \left[1 + (1+z_p)^{-\gamma-\kappa} \right]\,  \frac{(1+z)^{\gamma}}{1 + \left (  \frac{1+z}{1+z_p}   \right )^{\gamma + \kappa}}\, .
\ee
where $dV/dz$ is the differential comoving volume element and the $(1+z)$ factor at the denominator accounts for the source-frame to detector-frame time interval transformation.
The normalization factor of eq.\eqref{Madau} is the local BNS merger rate $R_0$, while the redshift $z_p$ is the peak of the star formation rate (which is in the range $z_p\sim 1.5-3$), and $\gamma$ and $\kappa$ are constants, defined so that this functional form  interpolates between a power-like behavior $R(z)\propto (1+z)^{\gamma}$ for redshifts well below $z_p$, and $R(z)\propto (1+z)^{-\kappa}$ at $z\gg z_p$. 

For the mass prior on the two BNS masses $m_1$ and $m_2$, we assume a joint flat distribution between $m_{\rm min}$ and $m_{\rm max}$ defined as
\be\label{massdistr}
p(m_1, m_2) dm_1 dm_2 = \dfrac{1}{\mathcal{N}} dm_1 dm_2 \, ,
\ee
where the normalization factor $\mathcal{N}$ is determined by requiring that $m_2 < m_1$. Specifically
\be
\begin{split}
    \mathcal{N} &= \int_{m_{\rm min}}^{m_{\rm max}} dm_1 \int_{m_{\rm min}}^{m_1} dm_2 \\
    &= \int_{m_{\rm min}}^{m_{\rm max}} \left(m_1 - m_{\rm min}\right) dm_1 \\
%    &= \dfrac{1}{2}\left(m_{\rm max}^2 - m_{\rm min}^2\right) - m_{\rm min}\left(m_{\rm max} - m_{\rm min}\right)\\
    &=\frac{1}{2}(m_{\rm max}-m_{\rm min})^2\, .
\end{split}
\ee
The population hyper-parameters, i.e. the parameters on which we will perform the inference, are therefore
\be\label{Lambdaz}
\Lambda_{\rm pop}=\{\gamma, \kappa, z_p, m_{\rm min}, m_{\rm max}\}\, .
\ee 
For cosmological inference, we consider two different scenarios. First, we consider the standard  (spatially flat) $\Lambda$CDM model. The expression of the luminosity distance as a function of redshift is then given by the usual expression
\begin{equation}\label{dLemmod}
d_L(z)=\frac{c}{H_0}\, (1+z) \,\int_0^z\, 
\frac{d\tilde{z}}{ 
\sqrt{\Omega_{\rm M} (1+\tilde{z})^3+ 
\Omega_{\rm R} (1+\tilde{z})^4
+ \ola }}\, ,
\end{equation}
where $\Omega_{\rm M}$  is the present matter  fraction, $\Omega_{\rm R}$ that of radiation (which we have written for completeness, but is completely negligible at the redshifts where astrophysical compact binaries merge), and $\ola$ is the energy density associated to a cosmological constant;  for a spatially flat cosmology,  $\Omega_{\rm M}+\Omega_{\rm R}+\ola=1$ . In this case, the cosmological hyper-parameters are
\be
\Lambda_{\rm cosmo}=\{H_0,\Omega_{\rm M}\}\, .
\ee
We then consider the modified gravity scenario. In this case, we fix $H_0$ and $\Omega_{\rm M}$ to their Planck~2018 values (this is essentially equivalent to the narrow $H_0$ prior used in \cite{LIGOScientific:2025jau}) and, in general,  the DE density becomes a function of redshift, $\ode (z)$; furthermore, the quantity extracted from the GW observation is the GW luminosity distance, related to the standard electromagnetic luminosity distance by 
\eqs{dLgwdLem}{eq:fit}, so the relation between the quantity $\dgw$ measured by GW observations and the redshift of the source is 
\be
\dgw(z)=\Xi (z)
\frac{c}{H_0}\, (1+z) \,\int_0^z\, 
\frac{d\tilde{z}}{ 
\sqrt{\Omega_{\rm M} (1+\tilde{z})^3+ 
\Omega_{\rm R} (1+\tilde{z})^4
+ \ode (\tilde{z}) }}\, .
\ee
For $\Xi(z)$ we adopt the parametrization (\ref{eq:fit}). For $\ode (z)$ a standard choice is the CPL parametrization~\cite{Chevallier:2000qy,Linder:2002et}
\be\label{4rdewdeproofs}
\ode(z)  =\Omega_{\rm DE}\exp\left\{ 3\int_{0}^z\, \frac{d\tilde{z}}{1+\tilde{z}}\, [1+\wde(\tilde{z})]\right\}\, ,
\ee
where $\Omega_{\rm DE}=\ode(z=0)$ and  
\be\label{w0wa}
w_{\rm DE}(z)= w_0+\frac{z}{1+z} w_a\, .
\ee 
With $H_0$ and $\Omega_{\rm M}$ fixed (and $\Omega_R$ fixed but in any case negligible), the cosmological hyper-parameters are therefore
\be
\Lambda_{\rm cosmo}=\{\Xi_0,n, w_0,w_a\}\, .
\ee
It should be observed, however, that (in a generic modified gravity theory, where these parameters are independent from those characterizing background evolution and scalar perturbations) the determination of $\Xi_0$ and $n$ can only be obtained with GW observations, while $\{w_0,w_a\}$, as well as
$\{H_0,\Omega_{\rm M}\}$, can be obtained from electromagnetic observations. In the following, we will assume that, by the time ET operates, narrow priors will be available on all quantities that can be determined electromagnetically, and we will use GW observation to infer $\Xi_0$ and $n$. Therefore, among the cosmological parameters, we will perform the inference only on them, and our set of cosmological parameters for the modified gravity case will be
\be
\Lambda_{\rm cosmo}=\{\Xi_0,n\}\, .
\ee
In particular, the cosmological parameters $H_0$ and $\Omega_{\rm M}$ are determined by a large number of electromagnetic observations,  so narrow priors are already available on them (and are expected to be even more stringent when 3G detectors will operate, in particular for $H_0$); of course,  performing an inference on $H_0,\Omega_{\rm M}$ from GW alone is interesting to see how much information GW observations can add to electromagnetic observations (in particular for $H_0$, in view of the Hubble tension, see \cite{H0DN:2025lyy} for a recent discussion). In contrast, performing an inference on $\{\Xi_0,n\}$ assuming that all we know on $H_0$ and $\oma$ comes from the BNS mass function itself is not very meaningful, and would artificially worsen the forecasts on $\{\Xi_0,n\}$.

\begin{table}[ht]
\centering
\begin{tabular}{ c c c }
\toprule
\midrule
\textbf{Parameter} & \textbf{Prior range (Uniform)} & \textbf{Fiducial value} \\ 
\midrule
$H_0$ & $[10, 200]$ km s$^{-1}$ Mpc$^{-1}$ & $67.66$ km s$^{-1}$ Mpc$^{-1}$\\
$\Omega_{\rm M}$ & $[0.05, 0.99]$ & $0.30966$\\
$\Xi_0$ & $[0.5, 2]$ & $1$ \\
$n$ & $[0.3, 5]$ & $1.91$ \\
$\gamma$ & $[0, 12]$ & $1.42$ \\
$\kappa$ & $[0, 6]$ & $4.62$ \\
$z_p$ & $[0, 4]$ & $1.84$ \\
$m_{\rm min}$ & $[0.1, 5] \, M_{\odot}$ & $1 \, M_{\odot}$ \\
$m_{\rm max}$ & $[0.1, 5] \, M_{\odot}$ & $2.5 \, M_{\odot}$ \\
\midrule
\bottomrule
\end{tabular}
\caption{Prior distributions and fiducial values adopted for the parameters reported in the leftmost column.}
\label{Table:priors}
\end{table}

The set of priors used on the population and cosmological data is shown in the middle column of
table~\ref{Table:priors}.
It is worth commenting on the prior on $n$ in the modified gravity case. As we have mentioned, the parametrization (\ref{eq:fit}) reduces to GR both for $\Xi_0=1$ (and arbitrary $n$) and  for $n=0$ (and arbitrary $\Xi_0$). 
If we only recovered GR for  $\Xi_0=1$ and arbitrary $n$, the fact that GR corresponds to a line in parameter space, rather than to a point, 
by itself would not be a problem; the consistency with GR, or the deviations from it,  would simply be quantified using the posterior for $\Xi_0$  marginalized over $n$. However, the existence of the GR line $n=0$ for arbitrary $\Xi_0$ complicates the picture. Furthermore,
this can create numerical problems in the numerical exploration of the parameter space, since for $n$ very close to zero the Markov Chain Monte Carlo  (MCMC) is led to explore arbitrarily large values of $\Xi_0$, resulting in unphysical tails in the posterior for $\Xi_0$; similarly, for $\Xi_0$ close to one, the MCMC could explore arbitrary values of $n$. These  problems are cured with a physically-motivated choice of prior for $n$.
The parametrization (\ref{eq:fit}) has been compared with the explicit behavior of $\dgw(z)/\dem(z)$ for a broad range of modified gravity models. Actually, it  was first  inspired by  non-local gravity, where it turns out that $n\sim 2-2.5$ (with precise values depending on the initial conditions of the model) in both the model of ref.~\cite{Maggiore:2013mea}, see \cite{Belgacem:2019lwx}, and 
in the model of ref.~\cite{Maggiore:2014sia}, see \cite{Belgacem:2018lbp}. The predictions for $n$ in a broad class of modified gravity models are shown in table~1 of 
\cite{LISACosmologyWorkingGroup:2019mwx}. While they generically depend on some free parameters, one sees from this table that typical values are not expected to be very close to zero (nor very large). For instance, in $f(R)$ gravity, adopting the  functional form for $f(R)$ proposed by Hu and  Sawicki~\cite{Hu:2007nk},  one gets (setting $\oma=0.3$) $n \simeq 0.3(\tilde{n}+1)$, where $\tilde{n}$ is a positive quantity, so $n\gsim 0.3$;  in ``designer $f(R)$ gravity'' \cite{Song:2006ej}, setting again $\oma=0.3$, one finds $n\simeq 2.3$; as another example, in models where the function $\delta(z)$ in \eq{prophmodgrav} follows the dark-energy density, 
$\delta(z)=\delta_0\ola(z)/\ola$, one finds $n=3\ola/\log (1/\oma)\simeq 1.7$;
see \cite{LISACosmologyWorkingGroup:2019mwx}. The bottomline is that, in typical models, $n$ is a number of order one, neither very close to zero nor very large. 
When performing a Bayesian analysis, we will then set a prior $n\in [n_{\rm min}, n_{\rm max}]$; we choose in particular $n_{\rm min}=0.3$ and $n_{\rm max}=5$. We find that the precise choice of $n_{\rm max}$ is not very important. In contrast, setting a prior $n\geq n_{\rm min}$  with $n_{\rm min}$ strictly positive  is important, otherwise we found that, in posterior of $\Xi_0$, large tails can be generated in the high-$\Xi_0$ region; in these cases we found that, even if the peak of the $\Xi_0$ posterior lies over the fiducial value,  the high-$\Xi_0$ tail introduces a significant offset in the median value, see appendix~\ref{app:priors} for details. We found that the value $n_{\rm min}=0.3$ is a good compromise between not restricting too much the parameter space, and ensuring numerical stability, and  that the results do  not change much taking a larger value of   $n_{\rm min}$ (while the error on $\Xi_0$ that we will present below might increase by a factor of order 2-3, when moving from $n_{\rm min}=0.3$ down to $n_{\rm min}=0$).    

\subsection{3G detector networks}

We consider ET both in isolation and in a network with CE. For ET we consider the two configurations currently under consideration by the ET Collaboration, namely a triangle with 10-km arms (denoted ``ET-$\Delta$''), and a configuration with two L-shaped detectors, each with 15~km arms, in two different sites (denoted ``ET-2L''). 

In the ET-2L setting, for definiteness the two L-shaped detectors will taken to be in the two candidate sites in Sardinia and in the Meuse-Rhine region, but no significant difference will appear with one L-shaped detector in Sardinia and the other in the candidate site near Kamenz, in the Lusatia region in Saxony. Indeed the great circle chord distance between the Sardinia site and the site in the Meuse-Rhine is 1165.0 km, while that between the Sardinia site  and Kamenz is 1247.3 km, so the two distances are quite comparable, with the Sardinia-Kamenz distance larger by about $7\%$. In contrast, the great circle chord distance between Kamenz and  the site in Meuse-Rhine is 575.4 km, significantly smaller, so all results concerning in particular angular localization would worsen significantly.
In the ET-2L configuration, the two L-shaped detectors are taken to be oriented at $45^{\circ}$ to each other,  where the relative orientation between  two L-shaped detectors is  defined with reference to the great circle that connects them~\cite{Flanagan:1993ix,Christensen:1996da}, similarly to what done in ref.~\cite{Branchesi:2023mws}, where a detailed comparison of the science output of these configurations was presented.\footnote{Actually, in ref.~\cite{Branchesi:2023mws} a small offset from $45^{\circ}$ was used in order not to send to zero the sensitivity to stochastic backgrounds (such small offset has little or no impact on parameter estimation for CBCs). Here we are only interested in CBCs, so we will not add such a small offset.} 

The sensitivity curve used for ET and for CE, as well as the location of the CE-40km detector, are also the same as in ref.~\cite{Branchesi:2023mws}.

\subsection{BNS event selection}

\begin{figure}[t]
    \centering
    \includegraphics[width=.6\textwidth]{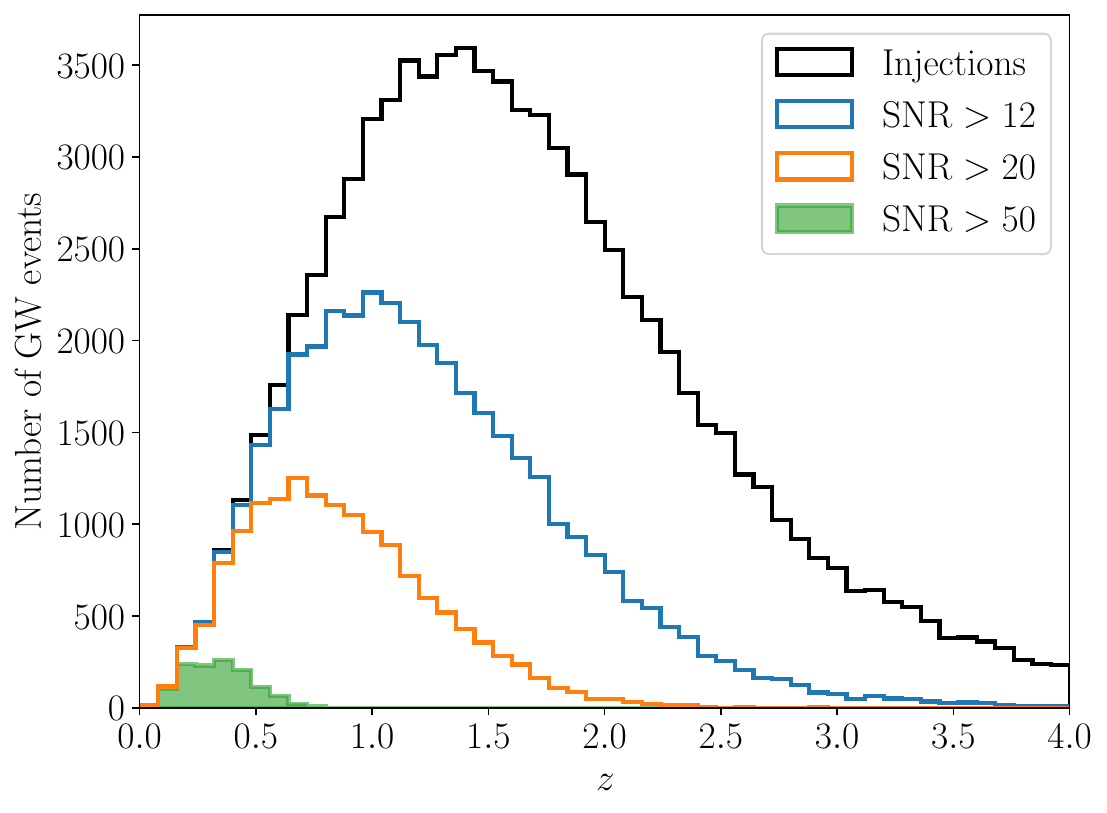}
    \caption{The distribution in redshift of the observed BNSs, for different cuts on the network SNR, for a network ET-2L+CE-40km. The black curve is the full injected BNS population, while the colored curves show the  distribution of the events recovered with a given SNR threshold.  In particular, the green-filled histogram represents the employed dataset.}
    \label{fig:redshift_distribution}
\end{figure}

We consider a BNS population described by the source parameters set 
\be
\boldsymbol{\theta} = \{\mathcal{M}, \eta, d_L, \chi_{1_{z}}, \chi_{2_{z}}, \theta, \phi, \iota, \psi, \Phi_{\rm coal}, t_{\rm coal}\}\, ,
\ee
where $\mathcal{M}$ is the chirp mass, $\eta$ is the symmetric mass ratio, $d_L$ is the luminosity distance, $\chi_{1_z}$ and $\chi_{2_z}$ are the dimensionless $z$-oriented spin components of the two compact objects, $\theta$ and $\phi$ are the colatitude and longitude, $\iota$ and $\psi$ are the inclination and polarization angle, while $\Phi_{\rm coal}$ and $t_{\rm coal}$ are the coalescence phase and time. For generating the BNS population, angles are generated assuming isotropy and uniformity. 
Distance and mass parameters are extracted from the redshift and individual mass distributions given in \eqs{Madau}{massdistr} with the appropriate transformations. The value of the fiducial parameters are shown in the rightmost column of table~\ref{Table:priors}. In particular, the Madau-Dickinson profile parameters $\gamma$, $\kappa$ and $z_p$ are taken from ref.~\cite{Iacovelli:2022bbs}, while $R_0$, $m_{\rm min}$ and $m_{\rm max}$ are chosen from the \textsc{Simple Uniform} BNS model\footnote{The BNS local merger rate is mass function dependent and ranges from $\mathcal{O}(10)$ to $\mathcal{O}(10^2)$. Following ref.~\cite{LIGOScientific:2025pvj}, results from the \textsc{FullPop-$4.0$} model feature a broad interval ($7.6-250$ Gpc$^{-3}$ yr$^{-1}$), while \textsc{Simple Uniform} yields tighter bounds ($13-170$ Gpc$^{-3}$ yr$^{-1}$). We choose to use the midpoint value of the latter range, as it matches our mass distribution assumptions and falls safely within both intervals.} reported in ref.~\cite{LIGOScientific:2025pvj}. In the injected population, we assume for simplicity the spin parameters to be identically zero, but we include them in the source parameter reconstruction of the signals. With these settings we obtain a total of $86049$ events per year.
Each event is then projected, assuming the \textsc{IMRPhenomD} waveform model \cite{Khan:2015jqa}, over the detectors' sensitivity curves, from $f_{\rm min} = 2$ Hz to from $f_{\rm max} = 2048$ Hz. We adopt a $100\%$ duty factor throughout the year and, due to the typical length of BNS signals at 3G detectors ($\mathcal{O}(1)$ day), the Earth's rotation is included. Fisher matrices  are then computed for the whole set of source parameters $\boldsymbol{\theta}$, resulting in a $11\times11$ matrix for each event. We recover posterior samples by the direct sampling of the high-SNR limit of the single-event posterior, defined as
\be
p(\boldsymbol{\theta}) d\boldsymbol{\theta} \propto \exp\left( -\dfrac{1}{2} \Delta\theta_i\Gamma_{ij}\Delta\theta_j \right) d\boldsymbol{\theta} \, ,
\ee
where $\Gamma$ represents the Fisher matrix. To avoid ill-conditioned results we exclude from this procedure all the events for which
\be
\delta_{\rm inv} = \max_{i,j} \left|\left(\Gamma\Sigma\right)_{ij} - \mathbb{I}_{ij} \right| > 5 \times 10^{-2} \, ,
\ee
where $\Sigma = \Gamma^{-1}$ and $\mathbb{I}$ is the identity matrix. The catalog generation relies on the python package \textsc{MGCosmoPop}~\cite{Mancarella:2021ecn}, while the whole Fisher-matrix-related approach mentioned above is performed through the python library \textsc{GWFAST}~\cite{Iacovelli:2022mbg}, to which we refer the reader for more details.

We decided to focus only on the BNSs for which the network signal-to-noise ratio (SNR) is relatively high, setting a threshold at ${\rm SNR}\geq 50$. This still gives $\mathcal{O} (10^3)$ BNSs/yr for a 3G network involving ET and CE, and $\mathcal{O}(10^2)$ BNSs/yr events for ET alone; this  is still a significant  sample, that allows us to perform a meaningful dark siren study. For a full 3G network with both ET and CE, the size of the sample could be increased by a factor about $\sim10$ ($\sim30$) lowering the SNR to $20$ ($12$). Analogously, for an ET-only observatory, the size would increase by a factor of about $\sim13$ ($\sim50$). However, we found that this introduces biases in the reconstruction of the injected values of the hyper-parameters. In particular, the poorer parameter estimation associated with low-SNR events can bias the exploration of parameter space toward incorrect regions, as the Fisher matrix approximation becomes increasingly unreliable in the low-SNR regime. Furthermore, our simulation does not take into account the effect on the parameter estimation due to the specific noise realization. While high-SNR events are not affected by this, a lower threshold makes the noise contribution less negligible and should be therefore taken into account properly.
Thus, to be on the safe side, we eventually restricted our analysis to the BNSs with ${\rm SNR}\geq 50$. This means that we are selecting BNSs at relatively low redshift. Figure~\ref{fig:redshift_distribution} shows the distribution in redshift of the observed BNSs, for different cuts on the network SNR, for a network ET-2L+CE-40km; we see that all BNSs that pass this cut are at $z<0.8$. On the one hand, the BNSs that do not pass this cut will have in general a worse parameter reconstruction, because they have a lower SNR, so in this sense one might think that they would (individually) contribute less to the determination of the cosmological parameters; on the other hand, modified GW propagation is a phenomenon that grows with redshift, and vanishes as $z\ra 0$, so the most distant objects are also those where the effect is more significant. Therefore, it is clear that restricting ourselves to BNSs with ${\rm SNR}\geq 50$ will provide a conservative  --- and possibly a very conservative --- forecast on the parameter $\Xi_0$.

\section{Results}\label{sect:results}
In this section we present the results obtained for the following networks: ET-$\Delta$, ET-2L, ET-$\Delta$+CE-40km and ET-2L+CE-40km. The SNR $= 50$ cut and the Fisher matrix inversion check leave, respectively, $163$, $378$, $962$ and $1237$ GW events to be employed in the hyper-parameter inference. We explore the parameter space using the \textsc{Icarogw} wrapper of the nested sampling algorithm \textsc{UltraNest} \cite{Buchner:2021cql}. The sampler is run with a standard evidence-based convergence criterion, controlled by the parameter \texttt{remainder\_fraction}. We adopt \texttt{remainder\_fraction} $ = 0.01$, such that the sampling is terminated once the estimated remaining contribution to the Bayesian evidence from the unexplored prior volume falls below 1\% of the evidence accumulated up to that point. This ensures that the evidence estimate has converged to the percent level.

\begin{table}[th]
\centering
\begin{tabular}{l c c c c c c c c c}
\toprule
\midrule
\textbf{Network} & $H_0$ & $\Omega_{\rm M}$ & $\gamma$ & $\kappa$ & $z_p$ & $m_{\rm min}$ & $m_{\rm max}$ & $z_{\rm best}$ & $H(z_{\rm best})$\\
\midrule
ET-$\Delta$ & $0.12$ & $1.02$ & $0.91$ & $0.45$ & $0.77$ & $0.017$ & $0.016$ & $0.23$ & $0.10$ \\
ET-2L & $0.11$ & $0.87$ & $0.93$ & $0.45$ & $0.80$ & $0.012$ & $0.014$ & $0.28$ & $0.06$ \\
ET-$\Delta$+CE-40km & $0.09$ & $0.47$ & $0.28$ & $0.47$ & $0.58$ & $0.010$ & $0.010$ & $0.37$ & $0.04$ \\
ET-2L+CE-40km & $0.09$ & $0.48$ & $0.26$ & $0.47$ & $0.64$ & $0.010$ & $0.011$ & $0.38$ & $0.03$ \\
\midrule
\bottomrule
\end{tabular}
\caption{Relative precision (as defined in footnote~\ref{foot:accuracy}) achieved for the cosmology and population hyper-parameters in $\Lambda$CDM cosmology, for the different network configurations. The columns labeled $z_{\rm best}$ and $H(z_{\rm best})$ refer respectively to the redshift at which $H(z)$ is best constrained and the relative precision on $H(z)$.}
\label{Table:precisions_H0_Om0}
\end{table}

\begin{table}[th]
\centering
\begin{tabular}{l c c c c c c c}
\toprule
\midrule
\textbf{Network} & $\Xi_0$ & $n$ & $\gamma$ & $\kappa$ & $z_p$ & $m_{\rm min}$ & $m_{\rm max}$ \\
\midrule
ET-$\Delta$ & $0.18$ & $0.84$ & $1.07$ & $0.44$ & $0.76$ & $0.013$ & $0.010$ \\
ET-2L & $0.18$ & $0.77$ & $1.00$ & $0.44$ & $0.80$ & $0.008$ & $0.008$ \\
ET-$\Delta$+CE-40km & $0.06$ & $0.87$ & $0.30$ & $0.46$ & $0.57$ & $0.005$ & $0.005$ \\
ET-2L+CE-40km & $0.06$ & $0.83$ & $0.30$ & $0.47$ & $0.62$ & $0.005$ & $0.004$ \\
\midrule
\bottomrule
\end{tabular}
\caption{Relative precision (as defined in footnote~\ref{foot:accuracy}) achieved for the cosmology and population hyper-parameters in modified gravity, for the different network configurations. $H_0$ and $\Omega_{\rm M}$ are now fixed to their fiducial values.}
\label{Table:precisions_Xi0_n}
\end{table}

\begin{figure}
    \centering
    \includegraphics[width=\linewidth]{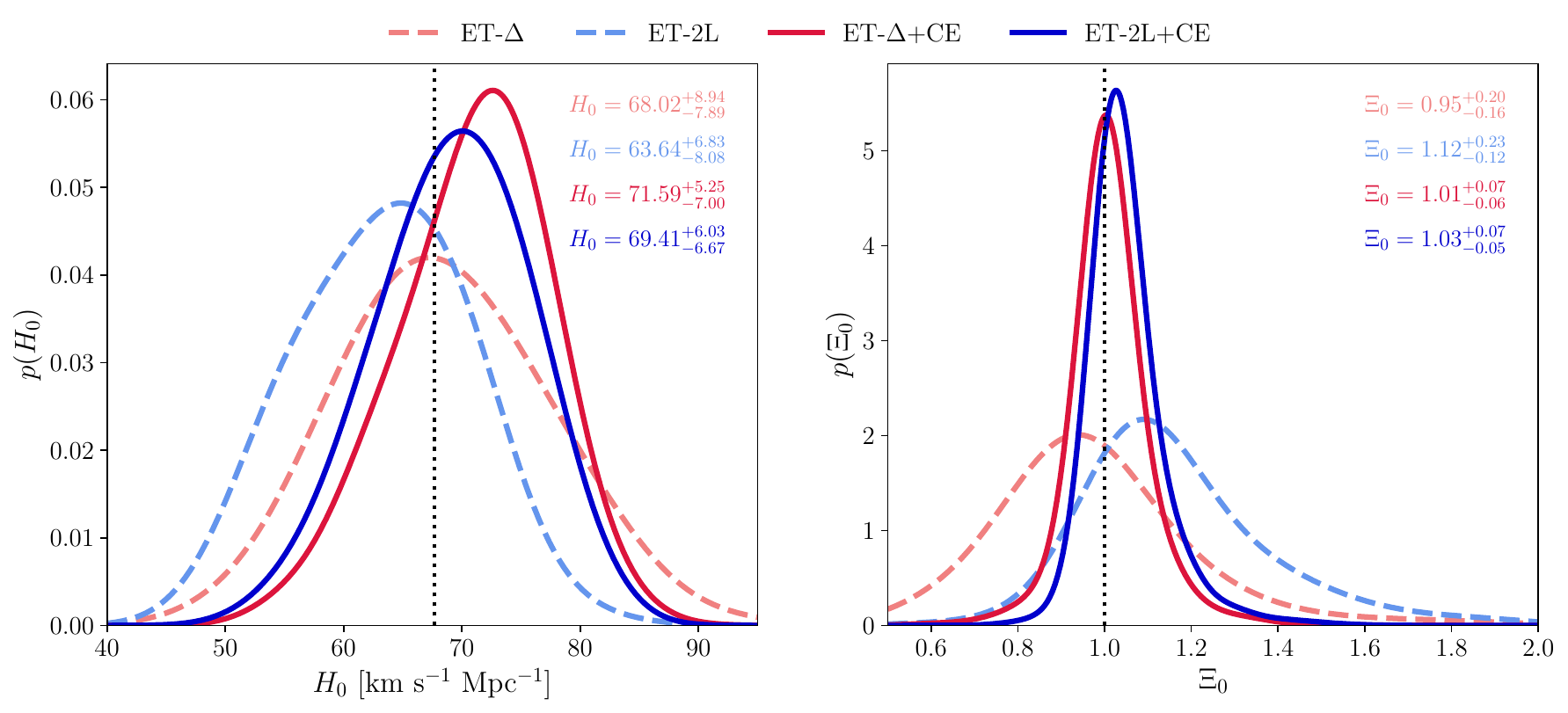}
    \caption{Posterior distributions for $H_0$ (left) and $\Xi_0$ (right) marginalized over the other hyper-parameters for the four networks considered. The black dotted line shows the fiducial values as referenced in table~\ref{Table:priors}. The posteriors are smoothed using a kernel distribution estimate.}
    \label{fig:H0_Xi0_summary_plots}
\end{figure}

Tables \ref{Table:precisions_H0_Om0} and \ref{Table:precisions_Xi0_n} summarize the relative precision with which parameters are measured, for the cases of $\Lambda$CDM and of modified gravity, respectively.\footnote{Given that the posteriors are in general not symmetric, we computed the relative precision 
 on each parameter in tables \ref{Table:precisions_H0_Om0} and \ref{Table:precisions_Xi0_n} as $(q_{84} - q_{16})/(2 \mu)$, where $q_n$ is the $n$-th quantile of the marginalized distributions and $\mu$ is the relevant fiducial value. This gives the $68\%$ c.l. and reduces to the the standard $1\sigma$ error for a symmetric posterior.\label{foot:accuracy}} Similarly, figure~\ref{fig:H0_Xi0_summary_plots} shows the posterior distributions for $H_0$ and $\Xi_0$, marginalized over the rest of the parameters, for all network configurations. 
A first remark is that both 3G networks perform similarly, regardless of the different number of events. 
Regarding $\Xi_0$, there is a significant improvement when CE is included in the network, whereas the change of ET configuration has no impact on the precision. This can be explained by the fact that the effect of modified gravity, as described by \eq{eq:fit}, goes to zero as $z\ra 0$ and grows monotonically with redshift, and the large redshift range is  more extensively probed by adding a CE detector.

\begin{figure}[h]
    \centering
    \includegraphics[width=0.49\linewidth]{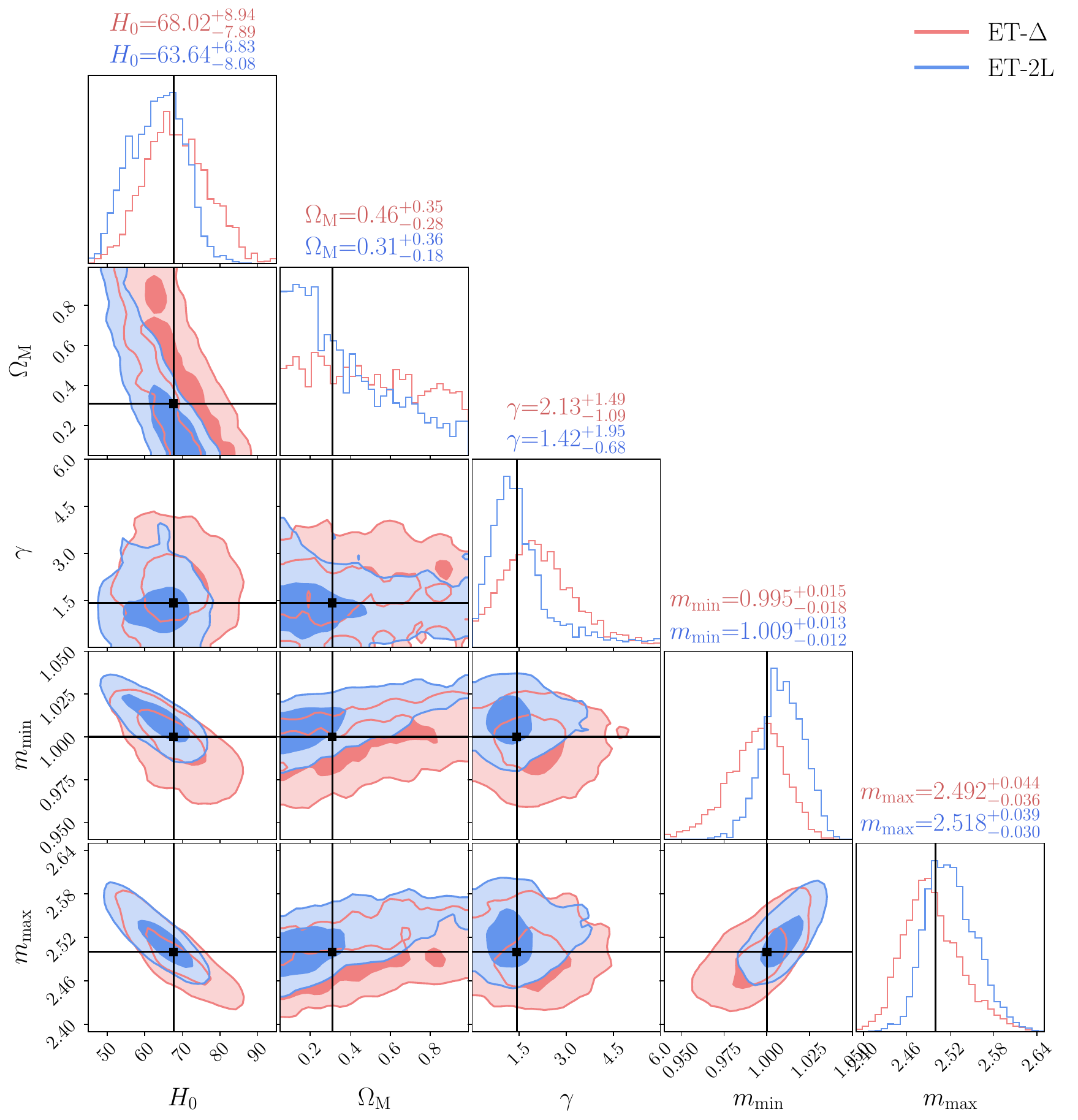}
    \includegraphics[width=0.49\linewidth]{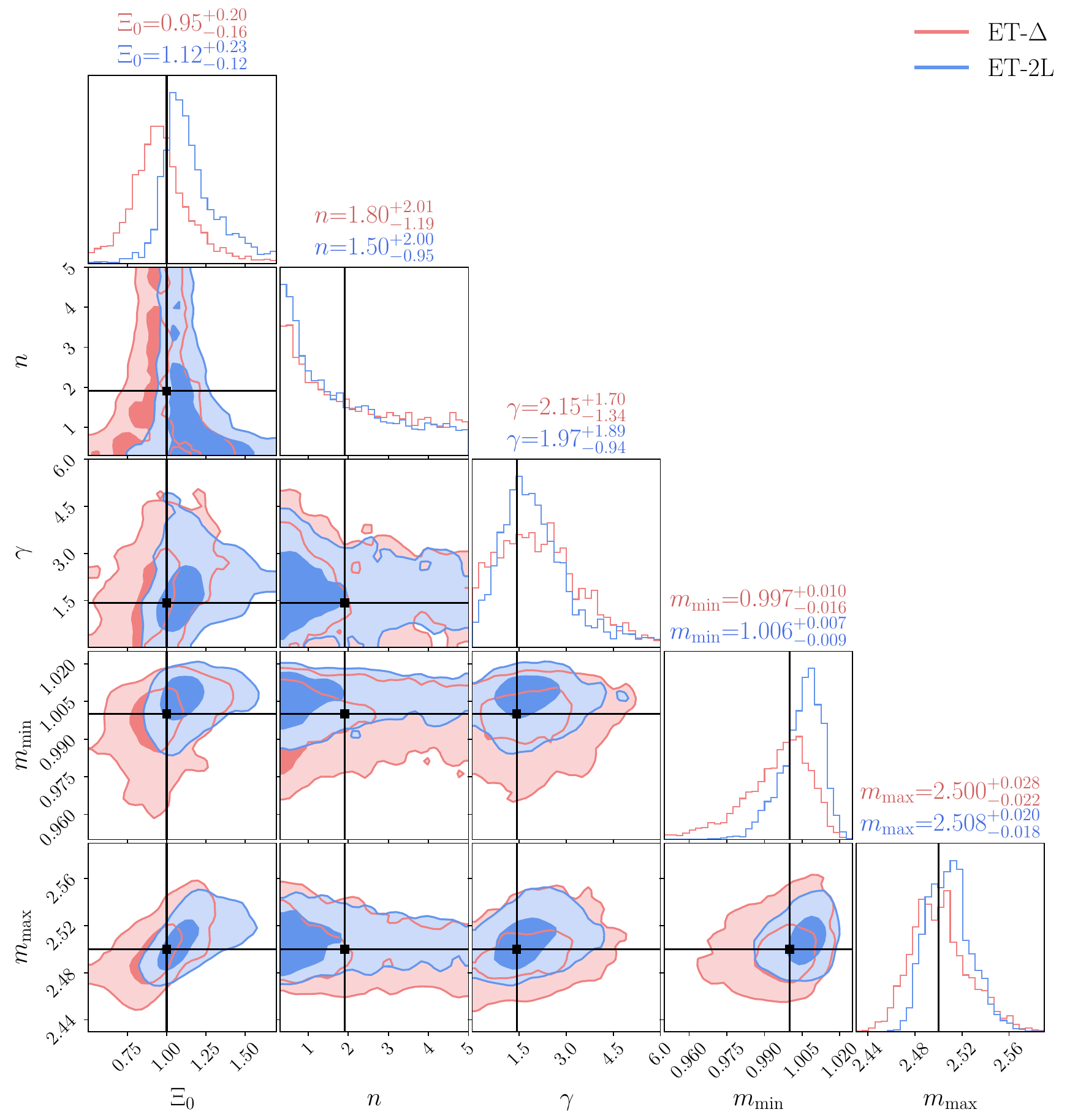}
    \caption{Posterior distributions of the parameters shown at the bottom for the ET-$\Delta$ and ET-2L network configurations. Contour shades represent, from dark to light, $1\sigma$ and $2\sigma$ credible regions. The black lines refer to the fiducial values. $\Xi_0$, $n$ (left) and $H_0$, $\Omega_{\rm M}$ (right) are fixed to their fiducial values.}
    \label{fig:results_ET}
\end{figure}

\begin{figure}[h]
    \centering
    \includegraphics[width=0.49\linewidth]{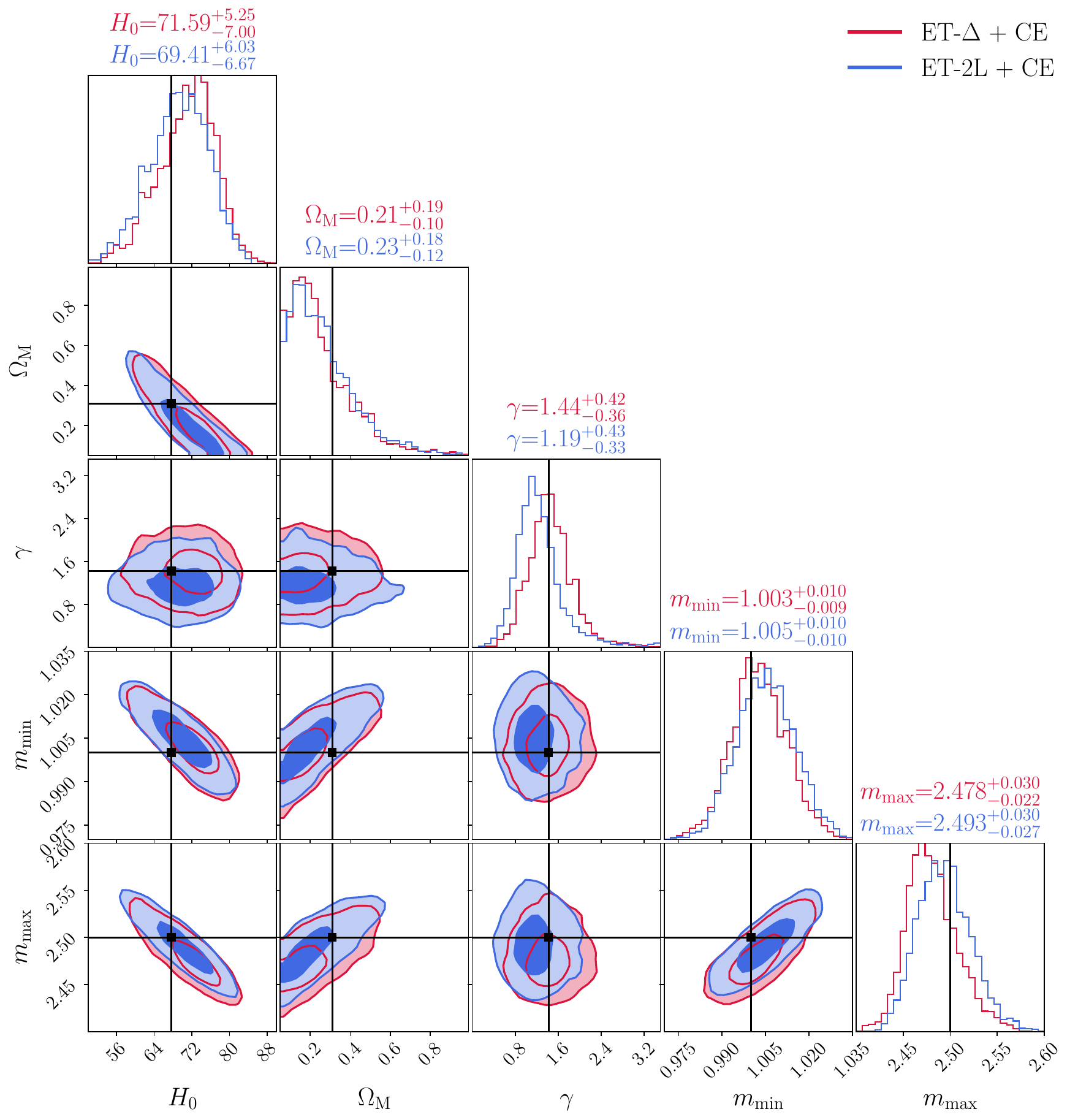}
    \includegraphics[width=0.49\linewidth]{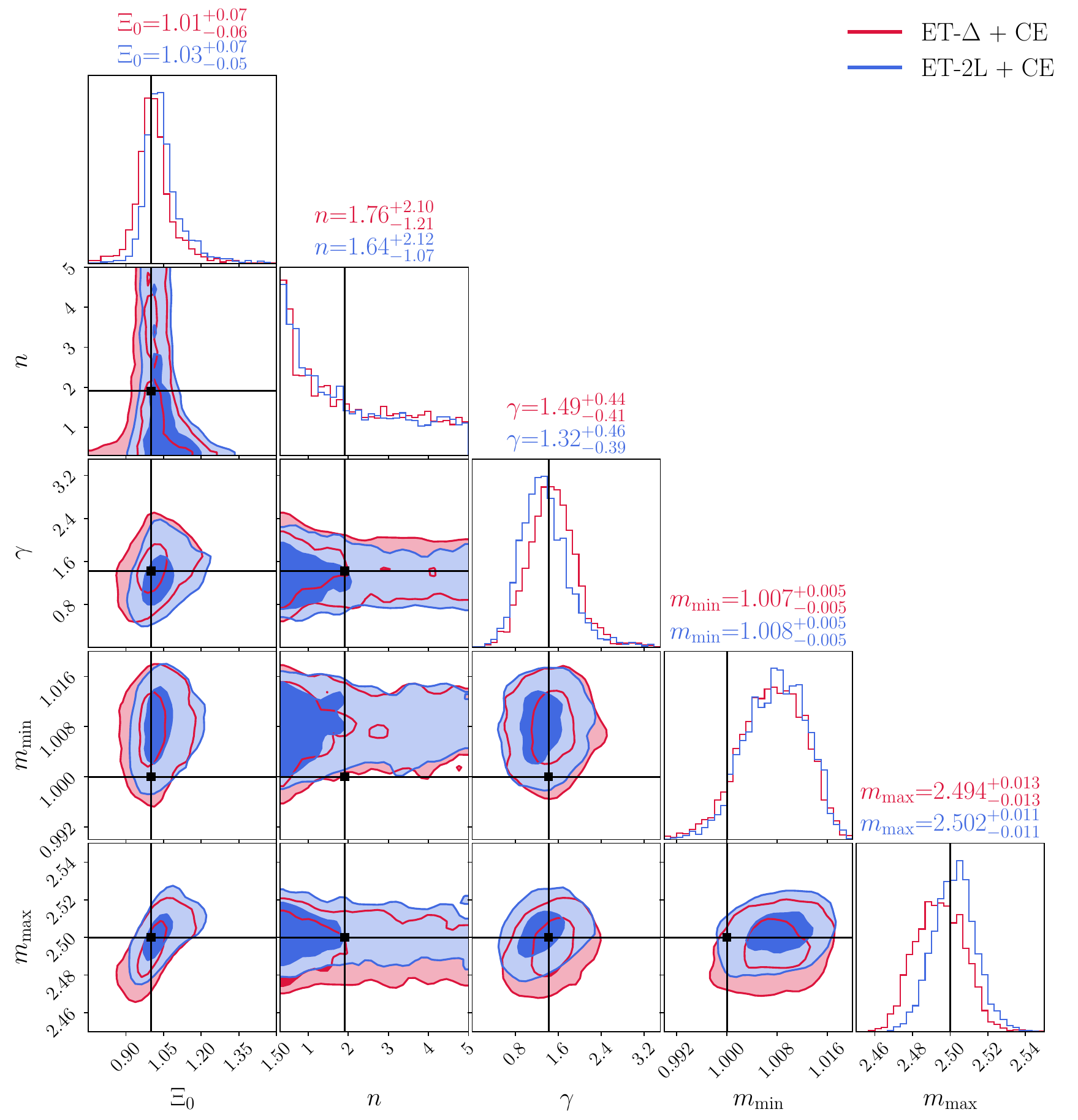}
    \caption{Posterior distribution of the parameters shown at the bottom for the ET-$\Delta$+CE and ET-2L+CE network configurations.  Contour shades represent, from dark to light, $1\sigma$ and $2\sigma$ credible regions. The black lines refer to the fiducial values. $\Xi_0$, $n$ (left) and $H_0$, $\Omega_{\rm M}$ (right) are fixed to their fiducial values.}
    \label{fig:results_ET+CE}
\end{figure}

Finally, figure~\ref{fig:results_ET} shows the results on the hyper-parameters for ET alone (triangle and 2L) and $\Lambda$CDM or modified gravity, while figure~\ref{fig:results_ET+CE} shows the analogous results when ET is in a network with CE. The constraints on $H_0$ are driven by its strong correlation with $\Omega_{\rm M}$ and the mass model parameters $m_{\rm min}$ and $m_{\rm max}$, as expected from the mass-redshift degeneracy.
Regarding the redshift evolution parameters, while $\gamma$ is well recovered owing to good coverage of our population in the redshift region $z<z_p$, $\kappa$ and $z_p$ are not well constrained due to a lack of events passing our cuts at redshifts above the Madau-Dickinson peak. For this reason, $\kappa$ and $z_p$ are marginalized over in figures \ref{fig:results_ET} and \ref{fig:results_ET+CE}. 

\section{Discussion}
\label{sect:conclusions}
Our results provide forecasts on the precision that can be expected from a small subset of BNSs, corresponding to the events with ${\rm SNR} > 50$, and therefore constitute only a lower bound on 3G capabilities. Indeed, for ET alone in the triangle (resp. 2L) configuration the selection criteria imposed in this study results in using  $1.8$\% (resp. $2.2$\%) of the events that pass the less stringent detection criterion ${\rm SNR} >12$; for ET-$\Delta$+CE (resp. ET-2L+CE) this fraction becomes  $2.8$\%  (resp. $3.2$\%).
Even if their SNR is lower,  we expect that the contribution of the events that have not been included can be significant, because of their large number compared to those that we have retained; furthermore, some of these events will be at higher redshift, where the effect of modified GW propagation is more important. Assuming a scaling of the relative precision with the number of events approximately as $1/\sqrt{N}$ (which can result from the fact that the events that we have not included have lower SNR but in general larger redshift and therefore better sensitivity to $\Xi_0$, as discussed above), we expect that a percent level accuracy on $\Xi_0$ could be reached.

\begin{figure}
    \centering
    \includegraphics[width=1.\linewidth]{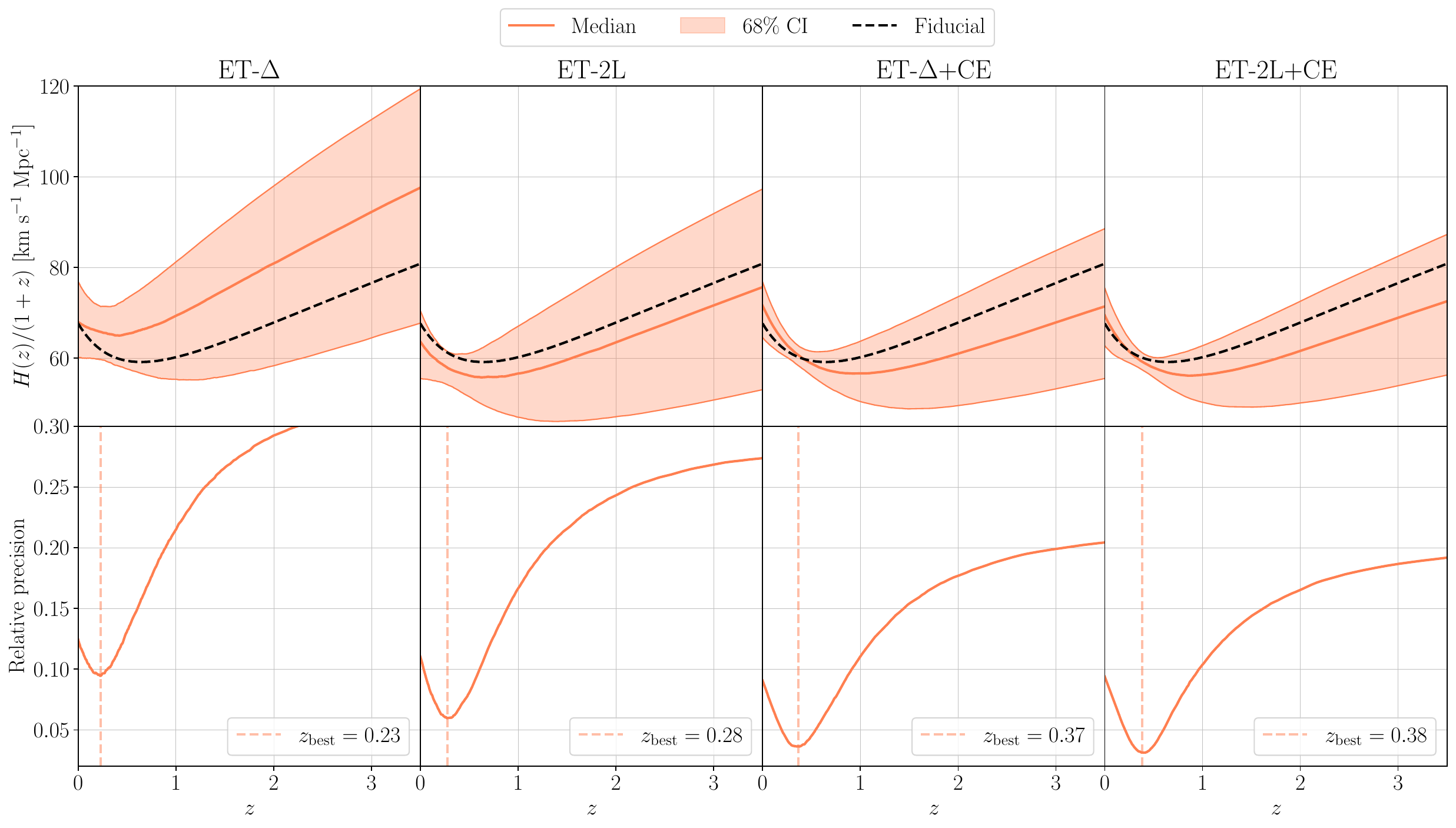}
    \caption{Constraints on the Hubble parameter $H(z)$ (top row) and its relative precision as defined in footnote~\ref{foot:accuracy} as a function of redshift (bottom row), for the configurations labeled at the top of each column. In the top row panels the shaded regions refer to the 68\% c.l., while the solid-orange (dotted-black) line labels the median (fiducial) curve.}
    \label{fig:H_of_z}
\end{figure}

Given the importance of events at high redshifts for modified GW propagation, another very interesting perspective would be to perform a similar study with BBHs, which are detectable at higher redshifts. Recent work has explored their potential as dark sirens to constrain the expansion history~\cite{Chen:2024gdn,Califano:2025qbx,Tagliazucchi:2026dpr}. 
In particular, there is consensus that, assuming a $\Lambda$CDM expansion history, BBHs as dark sirens provide their best constraint on $H(z)$ at intermediate redshifts rather than on $H_0$ and $\Omega_{\rm m,0}$ taken individually~\cite{Farr:2019twy,Mancarella:2021ecn,Farah:2024xub,Tagliazucchi:2026dpr}.
Specifically, for 3G detectors, ref.~\cite{Tagliazucchi:2026dpr} forecasts a 2.4\% precision on $H(z)$ at $z\sim1.5$ using $\sim 12000$ events at SNR$>60$. 
For comparison, our constraints on the $\Lambda$CDM expansion history correspond to a best constraint of $10\%$ ($6\%$) at $z\sim 0.23$ ($z\sim 0.28$) for the ET-$\Delta$ (2L) configuration when ET is alone, and a best constraint of $4\%$ ($3\%$) at $z\sim 0.37$ ($z\sim 0.38$) for the ET-$\Delta$ (2L) configuration in combination with CE, as illustrated by figure~\ref{fig:H_of_z}. 
These optimal redshifts are lower than the corresponding optimal redshift $z\sim1.5$ for BBHs~\cite{Tagliazucchi:2026dpr}, as a consequence of the lower SNR of BNSs. This suggests that the two samples can be combined to obtain a more precise joint measurement which takes advantage of both.  
For $\Xi_0$, on the other hand, there is no equivalent of the redshift that minimizes the relative uncertainty on deviations from GR. This is a consequence of the fact that the effect of modifications of gravity cumulates during propagation, in such a way that the parameter $\Xi_0$ encodes by design the maximum information that can be extracted from the data. 
In any case, eventually the optimal approach would be to work with a full CBC population to break degeneracies between the mass spectrum and the redshift distribution of the sources with a single hierarchical analysis~\cite{Ezquiaga:2022zkx,Chen:2024gdn}.
A crucial advantage of using BNSs, in particular, is that the BNS mass spectrum is restricted to a much narrower range, making its use less sensitive to the presence of putative peaks, changes in slope, and other features that can be prone to significant modeling systematics (see e.g.~\cite{Pierra:2023deu}); it is also expected to be more stable in redshift~\cite{Roy:2024oxh}.

Finally, an interesting extension of this work would be to jointly use the spectral and galaxy catalog methods. Indeed our BNS catalog covers $z<0.8$ which will be well-probed by upcoming galaxy surveys, such as Euclid \cite{Euclid:2021icp} and LSST \cite{LSST:2008ijt}. However, the galaxy catalog method, especially at the redshifts that can be reached by 3G detectors, must tackle delicate issues of catalog completeness (see \cite{Finke:2021aom,Borghi:2023opd,Dalang:2024gfk} for recent discussions). At the current level of knowledge, we preferred here not to mix methods or simultaneously combine different sets of assumptions and instead  study first their individual constraining power.

\newpage
%\vspace*{3mm}
\noindent
{\bf Acknowledgments.}
We thank Francesco Iacovelli, Andrea Ianniccari, Davide Perrone, Davide Piras for useful discussions.
This work is partially supported by the Spanish MCIN/AEI/
10.13039/501100011033 under the Grants No. PID2020-113701GB-I00, PID2023-146517NB-
I00 and CEX2024-001441-S, some of which include ERDF funds from the European Union, and by the MICINN with funding from the European Union NextGenerationEU (PRTR-C17.I1) and by the Generalitat de Catalunya. IFAE is partially funded by the CERCA program of the Generalitat de Catalunya. 
This project has received funding from the European Union’s Horizon Europe research and innovation programme under the Marie Skłodowska-Curie grant agreement No. 10181337.
The research of  M. Mag. and N.M. is supported by  the SwissMap National Center for Competence in Research. 
The research of M.~Mag. is supported by the SNSF grant CRSII5$\_$213497. 
The work of M.~Man. received support from the French government under the France 2030 investment plan, as part of the Initiative d'Excellence d'Aix-Marseille Universit\'e -- A*MIDEX AMX-22-CEI-02. 
Computations made use of the Baobab cluster at the University of Geneva. 

\appendix
\section*{Appendix A: Dependence of the choice of n priors}
\label{app:priors}

\begin{figure}[h!]
    \centering
    \begin{tikzpicture}
        \node at (0,0) {\includegraphics[width=\linewidth]{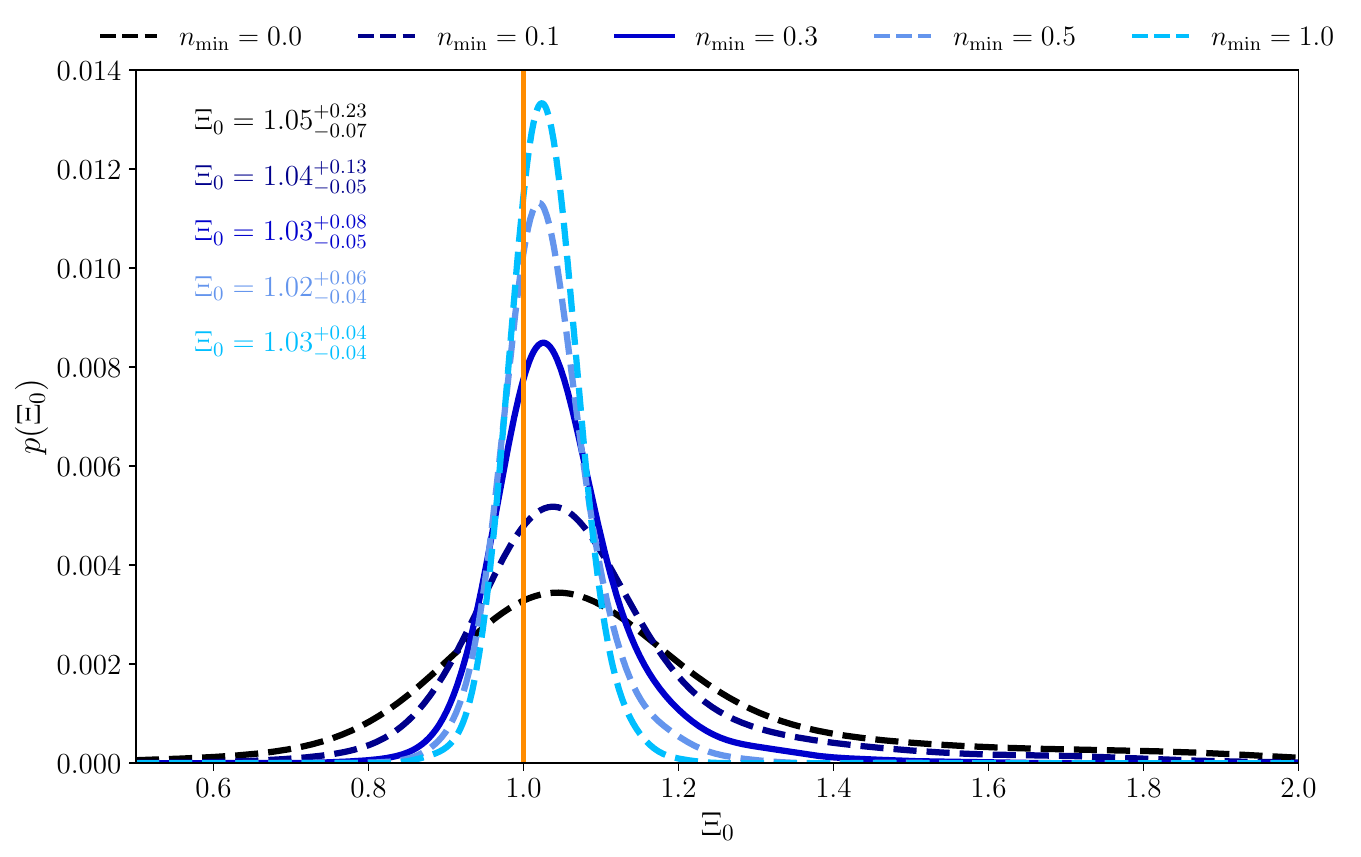}};
        \node at (4.3,1.3) {\includegraphics[width=0.35\linewidth]{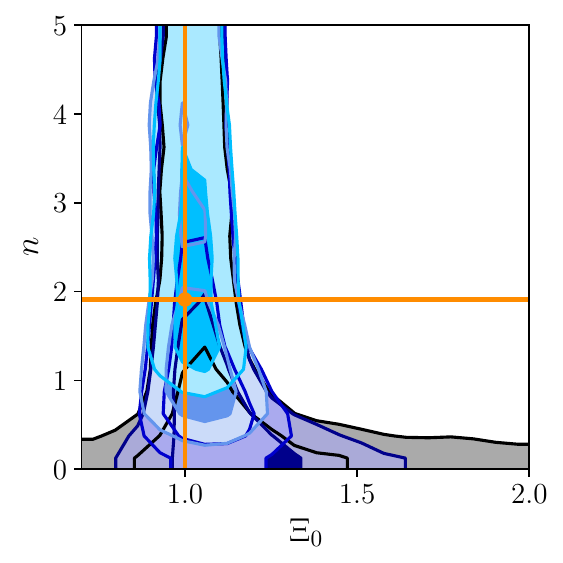}};
    \end{tikzpicture}
    \caption{Marginalized posteriors $p(\Xi_0)$ for different choices of the lower prior bound $n_{\rm min}$. Increasing $n_{\rm min}$ removes support at low $n$, tightening the distribution and shifting $\Xi_0$ toward $\simeq 1$. The inset shows the joint $p(\Xi_0,n)$ posterior, highlighting the parameter degeneracy and how truncating the prior in $n$ projects onto a narrower, shifted constraint on $\Xi_0$. The orange line refer to the fiducial value and the blue solid line corresponds to the final choice adopted in this study. Contour shades in the joint posterior distribution represent, from dark to light, $1\sigma$ and $2\sigma$ credible regions. $H_0$ and $\Omega_{\rm M}$ are fixed to their fiducial values, and the marginalization is performed over the remaining hyper-parameters.}
    \label{fig:n_priors}
\end{figure}

As discussed in section~\ref{sect:modGWprop}, there is a degeneracy between the modified gravity parameters $\Xi_0$ and $n$ for GR. Indeed, figure~\ref{fig:n_priors} illustrates that, when setting a wide priors, GR is compatible either with $\Xi_0=1$ for an arbitrary $n$ or $n=0$ for an arbitrary $\Xi_0$. We assess how the choice of $n_{\rm min}$ affects the marginal posterior distributions with the aim of identifying a physically motivated compromise to break this degeneracy. Setting $n_{\rm min}=0.3$ allows us to exclude the long tails in the $\Xi_0$ posterior, compatible with $n=0$, without narrowing our prior unnecessarily.

%\clearpage
\bibliographystyle{utphys}
\bibliography{bibliography.bib}

\end{document}

%% file: mydefs.tex
\newcommand{\calO}{\mathcal{O}}

\newcommand{\dgw}{d_L^{\,\rm gw}}
\newcommand{\dem}{d_L^{\,\rm em}}

\newcommand{\ra}{\rightarrow}

\def\lsim{\raise 0.4ex\hbox{$<$}\kern -0.8em\lower 0.62
ex\hbox{$\sim$}}

\def\gsim{\raise 0.4ex\hbox{$>$}\kern -0.8em\lower 0.62
ex\hbox{$\sim$}}

\def\lbar{{\hbox{$\lambda$}\kern -0.7em\raise 0.6ex
\hbox{$-$}}}

\newcommand\eq[1]{eq.~(\ref{#1})}
\newcommand\eqs[2]{eqs.~(\ref{#1}) and (\ref{#2})}

\newcommand\p{\partial}

\newcommand\ee{\end{equation}}
\newcommand\be{\begin{equation}}
\def\bea{\begin{array}}
\def\eea{\end{array}}\def\ea{\end{array}}
\newcommand\ees{\end{eqnarray}}
\newcommand\bees{\begin{eqnarray}}

\def\dslash{\hspace{-1mm}\not{\hbox{\kern-2pt $\partial$}}}
\def\Dslash{\not{\hbox{\kern-2pt $D$}}}
\def\pslash{\not{\hbox{\kern-2.1pt $p$}}}
\def\kslash{\not{\hbox{\kern-2.3pt $k$}}}
\def\qslash{\not{\hbox{\kern-2.3pt $q$}}}

\def\p1{{\bf p}_1}
\def\p2{{\bf p}_2}
\def\k1{{\bf k}_1}
\def\k2{{\bf k}_2}

\newcommand{\dddM}{\kern 0.2em \raise 1.9ex\hbox{$...$}\kern -1.0em \hbox{$M$}}
\newcommand{\dddQ}{\kern 0.2em \raise 1.9ex\hbox{$...$}\kern -1.0em \hbox{$Q$}}
\newcommand{\dddI}{\kern 0.2em \raise 1.9ex\hbox{$...$}\kern -1.0em\hbox{$I$}}
\newcommand{\dddJ}{\kern 0.2em \raise 1.9ex\hbox{$...$}\kern-1.0em
\hbox{$J$}}
\newcommand{\dddcalJ}{\kern 0.2em \raise 1.9ex\hbox{$...$}\kern-1.0em
\hbox{${\cal J}$}}

\newcommand{\dddO}{\kern 0.2em \raise 1.9ex\hbox{$...$}\kern -1.0em
\hbox{${\cal O}$}}
\def\dddz{\raise 1.5ex\hbox{$...$}\kern -0.8em \hbox{$z$}}
\def\dddd{\raise 1.8ex\hbox{$...$}\kern -0.8em \hbox{$d$}}
\def\dddbd{\raise 1.8ex\hbox{$...$}\kern -0.8em \hbox{${\bf d}$}}
\def\ddbd{\raise 1.8ex\hbox{$..$}\kern -0.8em \hbox{${\bf d}$}}
\def\dddx{\raise 1.6ex\hbox{$...$}\kern -0.8em \hbox{$x$}}
\newcommand{\ode}{\Omega_{\rm DE}}
\newcommand{\oma}{\Omega_{\rm M}}

\newcommand{\ola}{\Omega_{\Lambda}}

\newcommand{\wde}{w_{\rm DE}}